\documentclass[cits]{PoS}
\usepackage{graphicx}

\title{VLBI observations of helical jets: hints on the nature of radio-jets}

\ShortTitle{VLBI observations of helical jets.}

\author{\speaker{Manel Perucho}\\
        Universitat de Val\`encia, Dr. Moliner 50, 46100 Burjassot, Valencian Country, Spain\\
        E-mail: \email{manel.perucho@uv.es}}

\author{Yuri Y. Kovalev\\
        Astro Space Center of Lebedev Physical Institute, 117997 Moscow, Russia\\
        E-mail: \email{yyk@asc.rssi.ru}}
        
\author{Philip E. Hardee\\
        Department of Physics \& Astronomy, The University of Alabama, Tuscaloosa, AL 35487, USA\\
        E-mail: \email{phardee@bama.ua.edu}}

\author{Andrei P. Lobanov\\
        Max-Planck-Institut f\"ur Radioastronomie, Auf dem H\"ugel 69, 53121 Bonn, Germany\\
        E-mail: \email{alobanov@mpifr-bonn.mpg.de}}
        
\author{{Iv\'an Agudo}\\
        Instituto de Astrof\'{\i}sica de Andaluc\'{\i}a, Apartado 3004, 18080, Granada, Spain\\
        E-mail: \email{iagudo@iaa.es}}        
        
\author{Iv\'an Mart\'{\i}-Vidal\\
        Onsala Space Observatory (Chalmers University of Technology),
        Observatoriev\"agen 90, SE-43992 Onsala, Sweden\\  
        E-mail: \email{ivan.marti-vidal@chalmers.se}}

\abstract{We make use of VLBI observations of the radio jet in the quasar S5~0836+710 at different frequencies 
and epochs to study its properties. The jet shows helical structure at all frequencies. The ridge-line of the 
emission in the jet coincides at all frequencies and epochs, within the errors. We conclude that the helicity is  
a real, physical structure. Small differences between epochs reveal wave-like motion of 
the ridge-line transversal to the jet propagation axis. These transversal motions are measured to be superluminal. 
This unphysical result could correspond to a possible small amplitude oscillation of the ridge-line at the radio-core 
and to large errors in the determination of the positions. In addition, higher resolution images at 15~GHz show 
that the ridge-line does not coincide exactly with the centre of the radio jet. At arc-second scales, this powerful jet
shows non-collimated, irregular structure and a lack of a hot-spot. Following this collection of evidence, 
we conclude that the ridge-line could be related to a pressure maximum within the jet cross-section, associated 
with the observed helical pattern that could lead to jet disruption at longer scales.}

\FullConference{11th European VLBI Network Symposium \& Users Meeting,\\
		October 9-12, 2012\\
		Bordeaux, France}

\begin{document}

\section{Introduction}
Jets in active galactic nuclei (AGN) are observed mainly in the radio band, using the 
very long baseline interferometry (VLBI) technique. The nature and properties of the emitting region as related to 
the flow are still scarcely known. It is expected that the flow is subject to the
growth of different instabilities and that many of the structures observed (knots, bendings, helices) are caused by this physical process \cite{ha06,ha11,pe11}. We present here the conclusions derived from observations of the jet in S5~0836+710 at different frequencies and epochs showing that the 
ridge line of this jet behaves as expected if it is interpreted as a pressure wave \cite{pe+12a,pe+12b}.

The luminous quasar S5~0836+710 at a redshift $z=2.16$ hosts a powerful
radio jet extending up to kiloparsec scales \cite{hu92}. At this redshift,
$1\,\rm{mas} \simeq 8.4\,\rm{pc}$ (see, e.g., MOJAVE database). VLBI monitoring of the
source showed kink structures \cite{kr90} and yielded estimates of the bulk Lorentz factor
$\gamma_\mathrm{j}=12$ and the viewing angle $\alpha_\mathrm{j}=3^\circ$ of the
flow at milliarcsecond scales \cite{ot98}. The jet was observed
at 1.6 and 5~GHz with VSOP (VLBI Space Observatory Program, a Japanese-led 
space VLBI mission), and oscillations of the ridge-line were also shown \cite{lo98, lo06}.
It has been shown that the
presence of a shear layer allows fitting all the observed oscillation
wavelengths within a single set of parameters, assuming that they are produced
by KH instability growing along a cylindrical outflow \cite{pl07}. A relation between amplitude growth of the helical structure,
and a lack of a collimated jet structure and hot-spot where the jet interacts 
with the ambient medium was also verified \cite{pe+12a,pe+12b}. 

\section{Observations and analysis}   

Different observations at various frequencies and epochs were used for this work (see Table~1 in \cite{pe+12a}): VLBA and VSOP at 1.6 and 5~GHz \cite{lo06} at three and two different epochs,
respectively, two epochs at 1.6~GHz from EVN (European VLBI Network, the EVN is a joint facility of European, Chinese, South African and other
radio astronomy institutes funded by their national research councils), one including MERLIN \cite{pe+12b}, one epoch of simultaneous global VLBI (including VLBA) observations at 2 and 8~GHz (01/1997) \cite{pk12}, two epochs from VLBA at 8~GHz, three epochs from VLBA at 22 and 43~GHz, and 13 epochs, between 1998 and 2009, from the 2\,cm~VLBA/MOJAVE database at 15~GHz. 

The ridge-lines were calculated by determining the centre of the jet
emission (fitted by a Gaussian) at a given radial distance from the core. This was done radially outwards to obtain a complete picture of the ridge-line of the jet. We tested possible deviations in the computed ridge-line by obtaining the ridge-line
using two other approaches: from the location of the emission maximum and from the 
geometrical centre of the emission profile in a transversal slice above a certain image
rms cutoff level. These two different approaches are very similar to the former at low frequencies, where the jet 
is not resolved. However, at higher frequencies, the emission maximum does not necessarily coincide with the geometrical centre of the profile or the centre of the Gaussian (see Fig.~\ref{fig2}).

\begin{figure}
\begin{center}
\includegraphics[width=.6\textwidth]{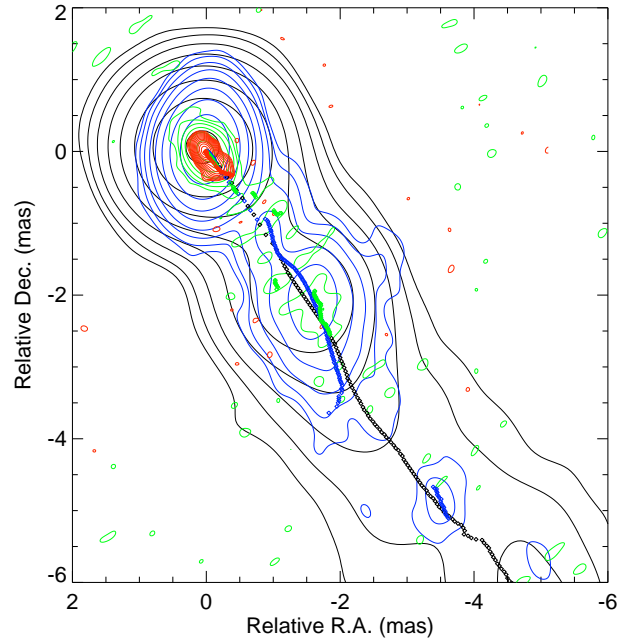}
\caption{Radio maps at the 8 (black contours), 15 (blue contours), 22 (green
contours) and 43~GHz (red contours) of the jet in 0836+710 in 1998. The position of the ridge-lines coincide within errors.}
\end{center}
\label{fig1}
\end{figure}

\begin{figure}
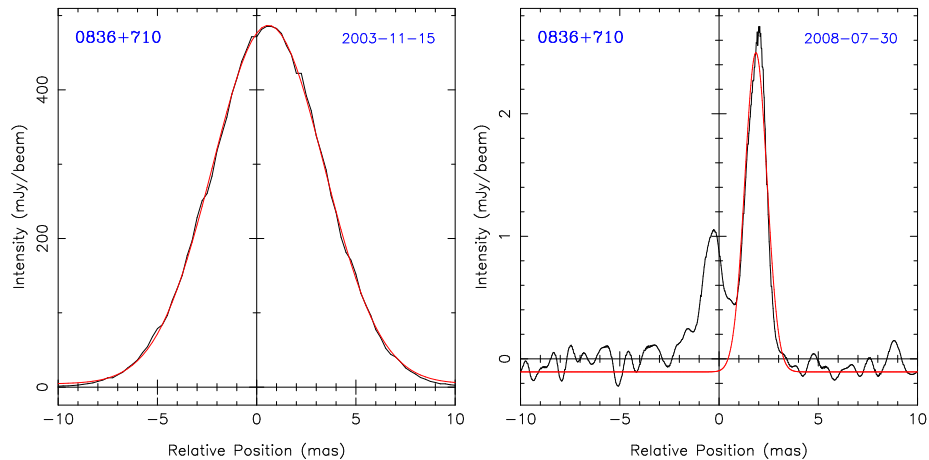

\begin{center}
\includegraphics[trim=0cm 0cm 0cm 1.6cm,clip,angle=0,width=0.4\textwidth]{fig2a.eps}
\includegraphics[trim=0cm 0cm 0cm 1.6cm,clip,angle=0,width=0.4\textwidth]{fig2b.eps}
\caption{Profiles and Gaussian fits for 1.6 (2003), and 15~GHz (2006).
In the left panel (1.6~GHz), the cut is done 7.50~mas from the core along the axis (PA=$-153.4^\circ$), 
the peak intensity is 482.5~mJy/bm, the peak relative position is 0.591~mas, and the FWHM of the Gaussian is 6.630~mas.  
In the right panel (15~GHz), the cut is done 8.20~mas from the core along the axis (PA=$-153.4^\circ$), 
the peak intensity is 2.6~mJy/bm, the primary peak relative position is 1.859~mas, and the FWHM of the Gaussian is 
1.31~mas. The two plots show the separation between the maximum of emission and the geometrical centre of the emitting region
when the resolution is increased.}
\end{center}
\label{fig2}
\end{figure}

\section{Discussion}

\begin{figure}
\begin{center}
\includegraphics[width=0.32\textwidth]{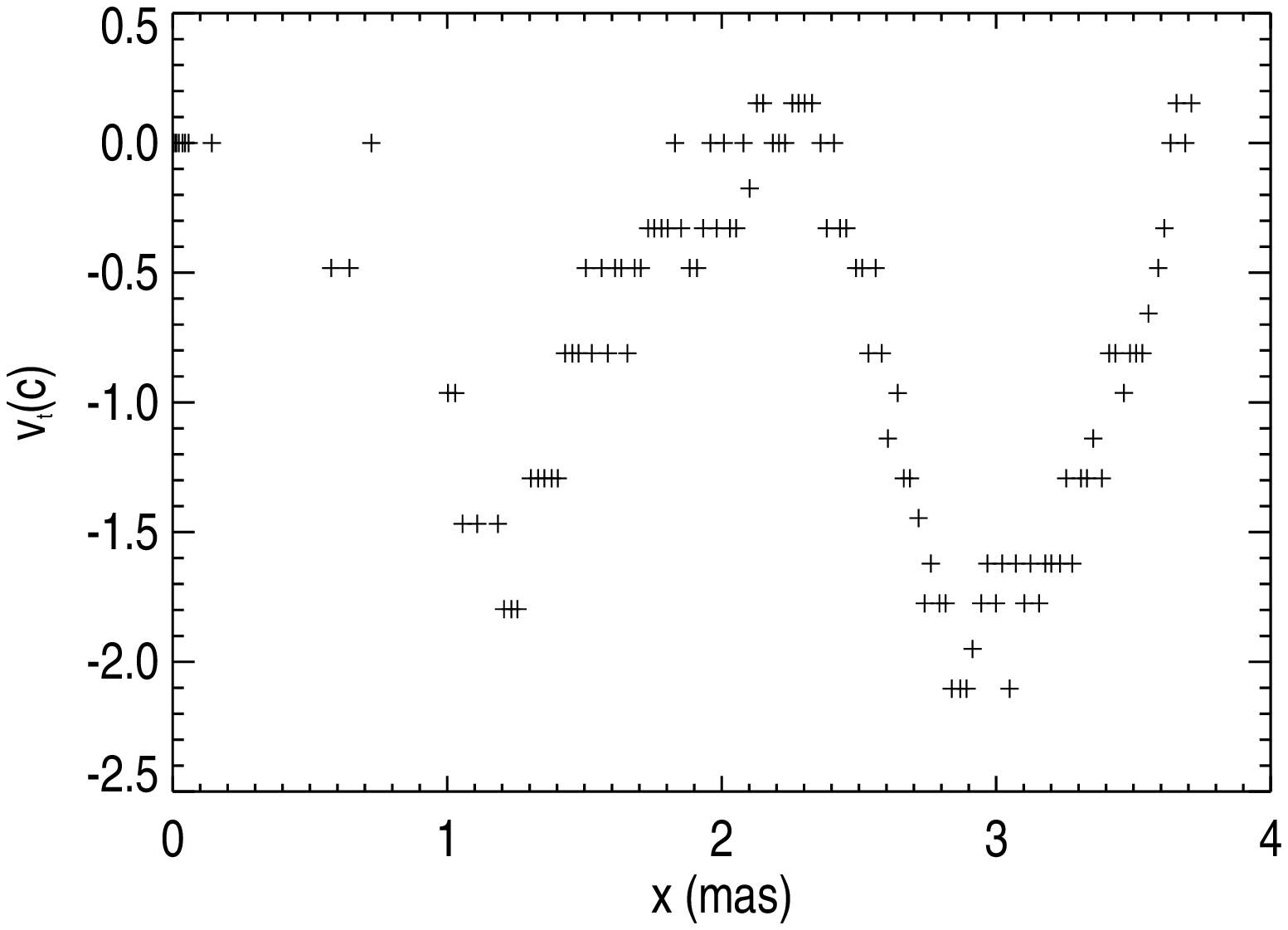}
\includegraphics[width=0.32\textwidth]{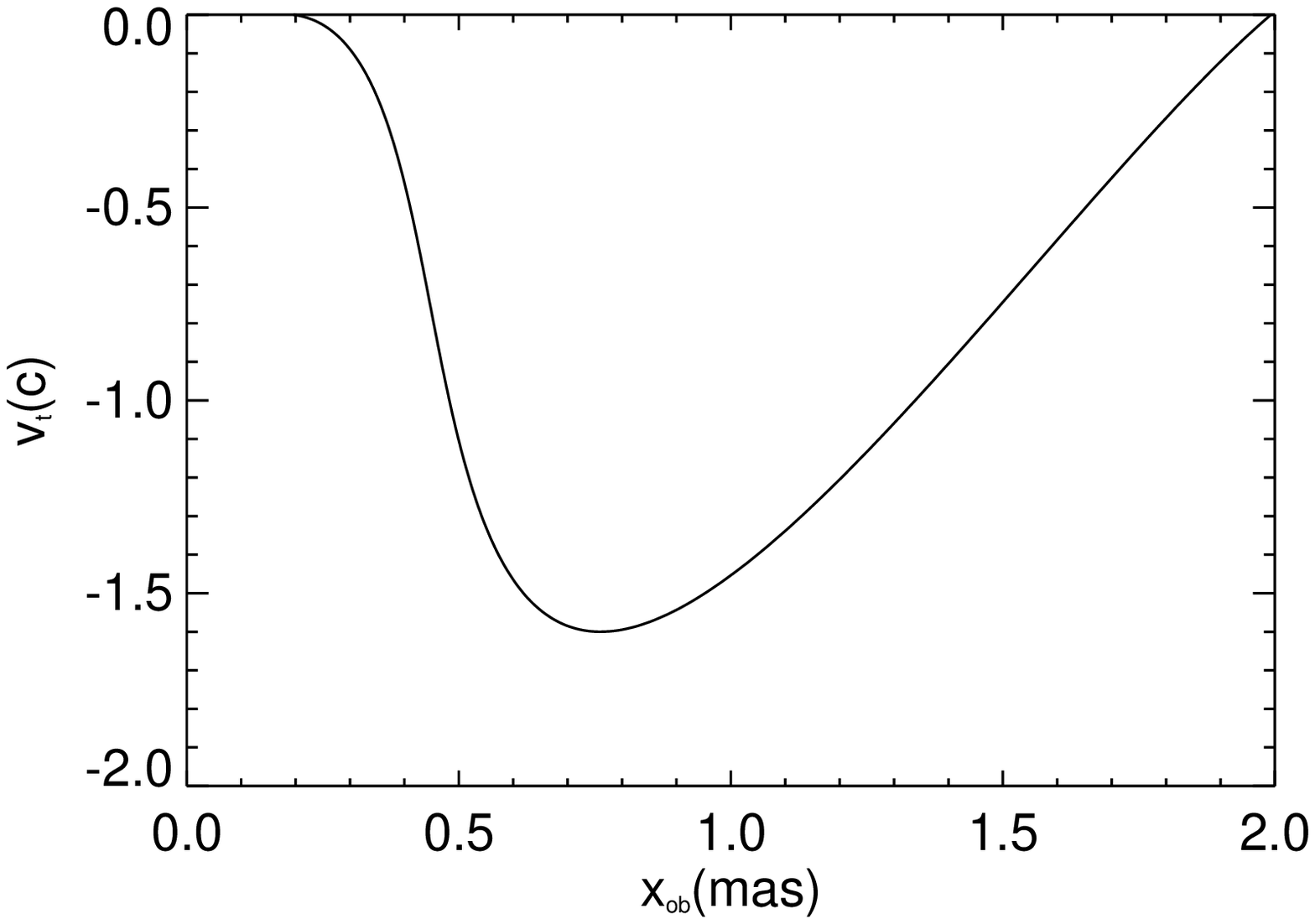}
\includegraphics[width=0.32\textwidth]{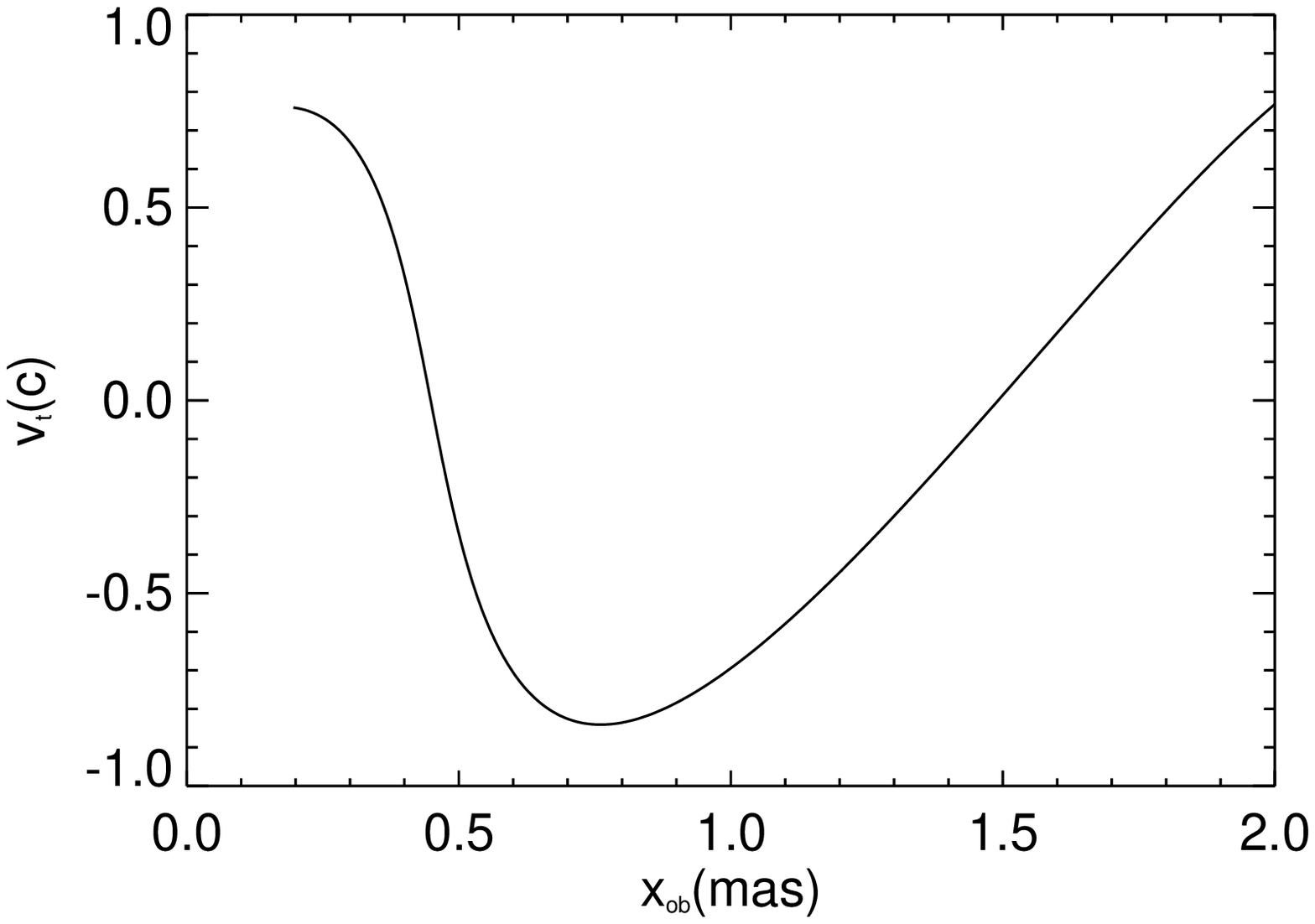}
\caption{Left panel: Transversal velocities versus the observed (projected) axial position, obtained from the transversal displacements measured at a fixed distance to the core, from 15~GHz VLBA data between two epochs in
2008 and 2009. Central panel: Same as left panel, for a simulated helical structure with relativistic speed ($0.96\,c$), at $3^\circ$ viewing angle and core position alignment. The two simulated epochs are separated by 2.11 years in the observer's reference frame. Right panel: Same as the central panel, without core position alignment. For details, see \cite{pe+12a}.}
\end{center}
\label{fig3}
\end{figure}

 The list of evidence collected from the ridge-line analysis \cite{pe+12a,pe+12b} can be summarized as follows: 1) The position of the ridge-line of the radio jet coincides at different frequencies. Different helical wavelengths show up 
 at different frequencies (Fig.~\ref{fig1}). 2) High resolution images confirm that this position does not necessarily coincide with the centre of emission of the radio jet (Fig.~\ref{fig2}). 3)The position angle of the radio jet changes with the frequency, as expected for a helical jet. 4) High resolution images also allow measurement of transversal displacements within the first mas, unaffected by relativistic effects, and show a clear wave-like oscillation pattern with distance. However, the obtained velocities are superluminal (left panel of Fig.~\ref{fig3}). 5) The opening angle measured for the jet at different frequencies is very similar and shows no correlation with frequency (the mean value is $12.1^\circ \pm 0.8$, which at a $3^\circ$ viewing angle implies an intrinsic opening angle of $0.63^\circ \pm 0.04$). 6) The amplitude of the oscillation grows along the jet propagation direction, as observed in the 1.6~GHz images. 7) At arc-second scales, the jet does not show any site of strong interaction with the ambient medium. This is interpreted as loss of collimation between hundreds of mas and arc-second scales.  
 
  As a result of the evidence, we conclude that the ridge-line corresponds to the pressure maximum, responsible for the helical structure of the radio jet. Moreover, the amplitude of the helix grows with distance and is possibly responsible for the loss of collimation. 
  
    A long-standing debate about jet physics was related to the nature of the radio jet: Is it flow that we are seeing or wave patterns? And, does the flow in the helical jets propagate non-ballistically? Our results show evidence for the radio jet being more related to a pattern in the flow. Thus, the flow could be propagating along the jet axis, and the radio jet would be revealing only a portion of the flow. If this is the case, the radio jet at high frequencies could be only a small part of the whole cross section, taking into account the changes of the jet position angle with frequency. Alternatively, the flow could be following the helical path, and this would not require the radio jet to be a small part of the flow cross section. The latter case is favored by the measurement of the opening angle, which is the same at all frequencies for which we have significant measurements, and the transversal velocities obtained from the observations at 15~GHz. The coincidence in the opening angles is less probable in the case of the radio jet corresponding to different regions across the jet. Thus, it seems that the radio emission is generated in the same region across the jet at all frequencies and that the jet flow indeed follows a helical path. A transition from ballistic to non-ballistic motion is compatible with the development of instabilities, which dissipate kinetic energy into internal energy. It has been reported that the development of the helical KH instability can force the jet flow into a helical path, however different from the helicity of the pressure maximum, the latter showing larger amplitude \cite{ha00}. The larger the Lorentz factor of the flow and the shorter the wavelength of the mode, the more different the helicity of the flow as compared to that of the pressure maximum. This can be understood in terms of the larger inertia of the jet flow with increasing Lorentz factor \cite{ha00,pe+05}. Within this picture, changes of the projection angle of the radio jet with time are expected, with the periodicities depending on the competing triggering frequencies of the modes, as revealed by the different observed wavelengths. Nevertheless, the transversal oscillation should be confirmed at 1.6~GHz to know whether the whole helix is oscillating, or it is only the smaller-scale ridge around the large-scale one.
  
  The transversal superluminal velocities can be understood in terms of 1) a small-scale displacement of the core, implying that the ridge-line at the core could also oscillate as it occurs downstream of the central engine (see central and right panels of Fig.~\ref{fig3}), and 2) the fact that errors in measuring the ridge-line position exceed the distance travelled by light ($0.012\,\rm{mas/yr}$) within the epochs that have been compared (1-2 years). It follows from our calculations, shown in Fig.~\ref{fig3}, that a wrong selection of the reference point can explain the superluminal transversal velocities on its own. This possible oscillation of the core position should be studied by phase-referencing experiments, and could have important implications regarding the nature of the core. In particular, this scenario implies that the core is not a special region in the jet and favors the view of it as a surface at which the jet becomes optically thin following a continuous process, rather than a discontinuous one, for instance, a reconfinement shock. 
 
  The pressure maximum associated with the ridge-line could couple to a growing instability, as indicated by the growth in amplitude of the helix with distance. It is difficult to discern between current-driven (CD) instability and KH instability modes, both being solutions to the linearized relativistic, magnetized flow equations and both being possible sources of helical patterns. Recent work on CD instability \cite{mi11} shows that a helical kink propagates with the jet flow if the velocity shear surface is outside the characteristic radius of the magnetic field. If the observed pattern corresponds to a CD kink instability, the observed transversal oscillation in the jet of 0836+710 requires that the kink be moving with the flow and implies that the transversal velocity profile is broader than the magnetic field profile, i.e., the
velocity shear surface lies outside the characteristic radius of the magnetic field. Thus, in this case the jet would have a magnetized spine surrounded by a particle dominated outer region. We would also like to point out that the development of the KH instability does not directly imply that the jet is particle dominated. It is however true that reasonable parameters for the jet result in KH growth rates in agreement with the observed growth in the amplitude of the helix. Further combined observational and theoretical studies like this one are required to try to get more information on the nature of the growing instability and on the properties of the jet flow.

\acknowledgments  
\small{MP acknowledges financial support by the Spanish ``Ministerio de Ciencia e Innovaci\'on''
(MICINN) grants AYA2010-21322-C03-01, AYA2010-21097-C03-01 and CONSOLIDER2007-00050. IA acknowledges funding by the ``Consejer\'{\i}a de Econom\'{\i}a, Innovaci\'on y Ciencia'' of the Regional Government of Andaluc\'{\i}a through grant P09-FQM-4784, an by the``Ministerio de Econom\'{\i}a y Competitividad'' of Spain through grant AYA2010-14844.}


\begin{thebibliography}{99}
\bibitem{ha06} P.E.~Hardee, \emph{AGN Jets: A Review of Stability and Structure}, in proceedings of \emph{Relativistic Jets: The Common Physics of AGN, Microquasars and Gamma-Ray Bursts}, eds.: P.A.~Hughes and J.N.~Bregman, \emph{AIP Conference Proceedings} {\bf 856} (2006) 57.
\bibitem{ha11} P.E.~Hardee, \emph{The stability of astrophysical jets}, in proceedings of \emph{IAU Symposium 275: Jets at all
Scale}, eds.: G.~Romero, R.~Sunyaev and T.~Belloni, \emph{IAU Conference Series} {\bf 275} (2011) 41.
\bibitem{pe11} M.~Perucho, \emph{Jet Stability, Dynamics and Energy Transport}, in proceedings of \emph{High Energy Phenomena
in Relativistic Outflows III} (HEPRO-III), eds. J.M. Paredes, M. Rib\'o, F.A. Aharonian and G.E. Romero, \emph{IJMPS} {\bf 8} (2012) 190. 
\bibitem{pe+12a} M.~Perucho, Y.Y.~Kovalev, A.P.~Lobanov, P.E.~Hardee, I.~Agudo, \emph{Anatomy of Helical Extragalactic Jets: The Case of S5~0836+710}, \emph{ApJ} {\bf 749} (2012) 55.
\bibitem{pe+12b} M.~Perucho, I.~Mart\'{\i}-Vidal, A.P.~Lobanov, P.E.~Hardee, \emph{S5~0836+710: An FRII jet disrupted by the growth of a helical instability?}, \emph{A\&A} {\bf 545} (2012) 65. 
\bibitem{hu92} C.A.~Hummel, T.W.B.~Muxlow, T.P.~Krichbaum, et al., \emph{MERLIN and VLBI observations of the quasar 0836+710: Morphology of a parsec-kiloparsec scale jet}, \emph{A\&A} {\bf 266} (1992) 93. 
\bibitem{kr90} T.P.~Krichbaum, C.A.~Hummel, A.~Quirrenbach, et al., \emph{The complex jet associated with the quasar 0836+71}, \emph{A\&A} {\bf 230} (1990) 271.
\bibitem{ot98} K.~Otterbein, T.P.~Krichbaum, A.~Kraus, et al., \emph{Gamma-ray to radio activity and ejection of a VLBI component in the jet of the S5-quasar 0836+710}, \emph{A\&A} {\bf 334} (1998) 489.
\bibitem{lo98} A.P.~Lobanov, T.P.~Krichbaum, A.~Witzel, et al., \emph{VSOP imaging of S5~0836+710: a close-up on plasma instabilities in the jet}, \emph{A\&A} {\bf 340} (1998) 60.
\bibitem{lo06} A.P.~Lobanov, T.P.~Krichbaum, A.~Witzel, J.A.~Zensus, \emph{Dual-Frequency VSOP Imaging of the Jet in S5~0836+710},Ê\emph{PASJ} {\bf 58} (2006) 253.
\bibitem{pl07} M.~Perucho, A.P.~Lobanov, \emph{Physical properties of the jet in 0836+710 revealed by its transversal structure}, \emph{A\&A} {\bf 469} (2007) L23.
\bibitem{pk12} A.B.~Pushkarev, Y.Y.~Kovalev, \emph{Single-epoch VLBI imaging study of bright active galactic nuclei at 2~GHz and 8~GHz}, \emph{A\&A} {\bf 544} (2012) 34.
\bibitem{ha00} P.E.~Hardee, \emph{On three-dimensional structures in relativistic hydrodynamic jets}, \emph{ApJ} {\bf 533} (2000) 176.
\bibitem{pe+05} M.~Perucho, J.M.~Mart\'{\i}, M.~Hanasz, \emph{Nonlinear stability of relativistic sheared planar jets}, \emph{A\&A} {\bf 443} (2005) 863.
\bibitem{mi11} Y.~Mizuno, P.E.~Hardee, K.-I.~Nishikawa, \emph{Three-dimensional relativistic magnetohydrodynamic simulations of current-driven instability with a sub-Alfv\'enic jet: Jet temporal properties}, \emph{ApJ} {\bf 734} (2011) 19.


\end{thebibliography}
\end{document}